\documentclass{eprint-ijmpa}

\begin{document}

\markboth{F.W. Stecker}
{High Energy Astrophysics Tests of Lorentz Invariance Violation}

%
\catchline{}{}{}{}{}
%

\title{HIGH ENERGY ASTROPHYSICS TESTS OF LORENTZ INVARIANCE VIOLATION\\
}

\author{\footnotesize F.W. STECKER}


\address{Laboratory for High Energy 
Astrophysics, NASA Goddard Space Flight Center\\
Greenbelt, MD, USA}

\maketitle


\begin{abstract}

Observations of the multi-TeV spectra of the Mkn 501 and other nearby BL Lac
objects exhibit the high energy cutoffs predicted to be the result of
intergalactic annihilation interactions, primarily with IR photons having 
a flux level as determined by various astronomical observations. After
correcting for such intergalactic absorption, these spectra can be explained
within the framework of synchrotron self-Compton emission models. Stecker and
Glashow have shown that the existence of this annihilation via 
electron-positron pair production puts strong constraints on Lorentz invariance
violation. Such constraints have important implications for some quantum 
gravity and large extra dimension models. A much smaller amount of Lorentz
invariance violation has potential implications for understanding the
spectra of ultrahigh energy cosmic rays.

\end{abstract}

\section{Consequences of Breaking of Lorentz Invariance} 

It has been suggested that Lorentz invariance (LI) may be only an approximate 
symmetry of nature \cite{sa72}. Although no true quantum theory of gravity 
exists, it was independently proposed that LI might be violated in such a 
theory with astrophysical consequences \cite{ac98}. A simple formulation
for breaking LI by a small first order perturbation in the electromagnetic 
Lagrangian which leads to a renormalizable treatment has been given by
Coleman and Glashow \cite{cg99}. Using this formalism, these authors 
point out that with LI violation (LIV), different particles can have maximum 
attainable velocities (MAVs) which can be different from $c$.
Using the formalism of Ref. \refcite{cg99}, we denote
maximum attainable 
velocity (MAV) of a particle of type $i$ (not necessarily equal to 
$c \equiv 1$) by $c_{i}$. We futher define the difference $c_{i} - c_{j}
\equiv \delta_{ij}$ and specifically here $c_{e\gamma} \equiv \delta << 1$. 
These definitions will be used to discuss the physics implications of
cosmic ray and cosmic $\gamma$-ray 
observations\cite{sg01},\cite{st03},\cite{st04}. 

If $\delta < 0$, 
the decay of a photon into an electron-positron pair is kinematically allowed
for photons with energies exceeding $E_{\rm max}= m_e\,\sqrt{2/|\delta|}$.
This decay would take place rapidly, so that photons with energies 
exceeding $E_{\rm max}$ could not be observed either in the laboratory or as 
cosmic rays. Since photons have been observed with energies   
$E_{\gamma} \ge$ 50~TeV from the Crab nebula \cite{ta98}, this implies 
that $E_{\rm max}\ge 50\;$TeV, or that $|\delta| < 2\times  
10^{-16}$.
 
If, on the other hand, $\delta > 0$, electrons become 
superluminal if their energies exceed $E_{\rm max}/\sqrt{2}$.
Electrons traveling faster than light will emit light  at all frequencies by a
process of `vacuum \v{C}erenkov radiation.' The
electrons then would rapidly lose energy until they become subluminal.
Because electrons have been seen in the cosmic radiation 
with energies up to $\sim\,$2~TeV\cite{ni00}, it follows that 
$\delta <  3 \times 10^{-14}$.
A smaller, but more indirect, upper limit on $\delta$ for the $\delta > 0$
case can be obtained from theoretical considerations of $\gamma$-ray emission 
from the Crab Nebula. Its emission above 0.1 GeV, extending into the TeV range, 
is thought to be Compton emission of the same relativistic electrons which
produce its synchrotron radiation at lower energies\cite{dh92}. The Compton component, 
extends to 50 TeV and thus implies the existence of electrons having energies 
at least this great
in order to produce 50 TeV photons, even in the extreme Klein-Nishina limit.
This indirect argument, based on the reasonable assumption that the 50 TeV
$\gamma$-rays are from Compton interactions, leads to a smaller upper limit on 
$\delta$, {\it viz.,} $\delta < 10^{-16}$.
~
A further constraint on $\delta$ for $\delta > 0$ 
follows from a change in the threshold energy for the pair 
production process $\gamma + \gamma \rightarrow e^+ + e^-$. 
This follows from the fact that the square of the 
four-momentum is changed to give the threshold condition

$$2\epsilon E_{\gamma}(1-cos\theta)~ -~ 2E_{\gamma}^2\delta ~\ge~ 4m_{e}^2,$$

\noindent where $\epsilon$ is the energy of the low energy photon and 
$\theta$ is the
angle between the two photons. The second term on the left-hand-side comes
from the fact that $c_{\gamma} = \partial E_{\gamma}/\partial p_{\gamma}$.
It follows that the condition for a significant increase in the energy
threshold for pair production is $E_{\gamma}\delta/2$ $ \ge$
$ m_{e}^2/E_{\gamma}$, or 
equivalently, $\delta \ge {2m_{e}^{2}/E_{\gamma}^{2}}$.
~
The $\gamma$-ray spectrum of the active galaxy Mkn 501 while flaring  
extended to $E_{\gamma} \ge 24$ TeV \cite{ah01} and exhibited the 
high energy absorption expected from $\gamma$-ray annihilation by extragalactic pair-production interactions with extragalactic infrared 
photons\cite{ds02}, \cite{ko03}.
This has led Stecker and Glashow \cite{sg01} to point out that the Mkn 501 
spectrum presents evidence for pair-production with no indication of 
Lorentz invariance violation (LIV) up to a photon energy of 
$\sim\,$20~TeV and to thereby place a quantitative constraint on LIV
given by $\delta < 2m_{e}^{2}/E_{\gamma}^{2} \simeq 
10^{-15}$. This constraint on positive $\delta$ is 
more secure than the smaller, but indirect, limit given above.
~
~
\section{Constraints on Quantum Gravity and Extra Dimension Models}

LIV has been proposed consequence of quantum gravity physics at the Planck 
scale $M_{Planck} = \sqrt{\hbar c/G} \simeq 1.22 \times 10^{19}$ GeV, 
\cite{ga95}, \cite{al02}. In models involving large extra 
dimensions, the energy scale at which gravity becomes strong can occur 
at a scale, $M_{QG} << M_{Planck}$, even 
approaching a TeV \cite{el01}.
In the most commonly considered case, the usual relativistic 
dispersion relations between energy and momentum of the photon and the electron
are modified\cite{ac98}, \cite{al02} by a term of order 
$p^3/M_{QG}$.\footnote{We note that there are
variants of quantum gravity and large extra dimension
models which do not violate Lorentz invariance and for which the constraints
considered here do not apply. There are also variants for which there
are no cubic terms in momentum, but rather much smaller quartic terms of order 
$\sim {p{^4}/ M_{QG}^2}$.}

Generalizing the LIV parameter $\delta$ to an energy dependent form

\begin{equation}
\delta~ \equiv~ {\partial E_{e}\over{\partial p_{e}}}~ -~ {\partial E_{\gamma}
\over{\partial p_{\gamma}}}~
 \simeq~ {E_{\gamma}\over{M_{QG}}}~ 
-~{m_{e}^{2}\over{2E_{e}^{2}}}~ -~ {E_{e}\over{M_{QG}}} ,
\end{equation}

\noindent the threshold condition from pair production implies
$M_{QG} ~\ge~ E_{\gamma}^3/8m_{e}^2.$
Since pair production occurs for energies of at least 20 TeV,
we find a constraint on the quantum gravity scale\cite{st03}
$M_{QG} \ge 0.3 M_{Planck}$. 
This constraint contradicts the 
predictions of some proposed quantum gravity models involving large extra 
dimensions and smaller effective Planck masses. In a variant model of Ref.
\refcite{el04}, the photon dispersion relation is changed, but not that of the 
electrons. In this case, we find the even stronger constraint $M_{QG} \ge 0.6 
M_{Planck}$. 

Within the context of a more general cubic modification of the dispersion 
relations, Jacobson, {\it et al.}\cite{jlm03} 
obtained an indirect limit on $M_{QG}$ from the apparent cutoff in the 
synchrotron component of the in the Crab Nebula $\gamma$-ray emission at 
$\sim 0.1$ GeV. However, their very strong constraint, 
$M_{QG}$$ >$$ 1.2 \times 10^{7} M_{Planck}$,
is qualified by considerations of electron helicity and photon 
polarization\cite{ja04}
and is thus not as general as the constraint from photon-photon 
pair-production. Also, for the model suggested in
\refcite{el04}, this constraint does not hold.

\section{LIV and the Ultrahigh Energy Cosmic Ray Spectrum}

Coleman and Glashow \cite{cg99} have shown that for 
interactions of protons with CBR photons of energy $\epsilon$ and 
temperature $T_{CBR} = 2.73 K$, pion production is kinematically forbidden and 
thus {\it photomeson interactions are turned off} if

$$\delta_{p\pi} > 5 \times 10^{-24}(\epsilon/T_{CBR})^2.$$

\begin{figure}
\vspace{-1.5cm}
\centerline{\psfig{figure=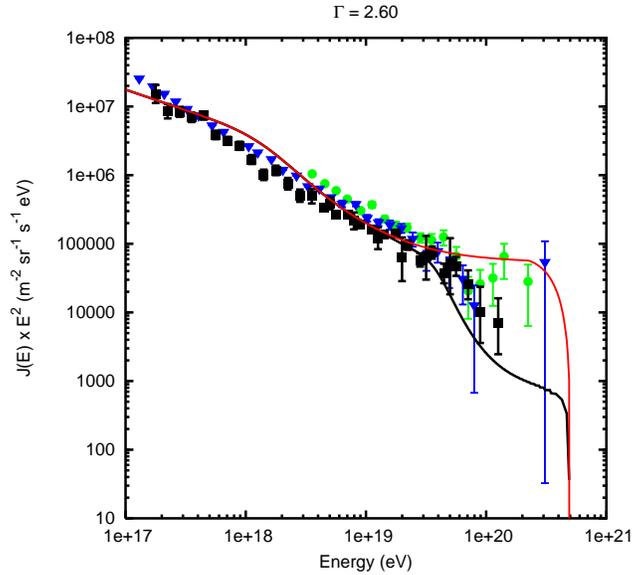,height=12cm}}
\vspace{-1.5cm}
\caption{Predicted spectra for an $E^{-2.6}$ source spectrum with 
redshift evolution and $E_{max}$ = 500 EeV, 
shown with pair-production losses 
included and photomeson 
losses both included (black curve) and turned off (lighter (red) curve). 
The curves are shown with ultrahigh energy cosmic ray spectral data from
{\it Fly's Eye} (triangles), {\it AGASA} (circles), 
and {\it HiRes} monocular data (squares)\protect\cite{st04}.}
\label{f2}
\end{figure}

Thus, given even a very small amount of LIV, photomeson and 
pair-production interactions of UHECR with the CBR can be turned off. 
Such a violation of Lorentz invariance might be produced by
Planck scale effects\cite{al00}, \cite{ap03}.
If Lorentz invariance violation is the explanation for the missing GZK effect,
indicated in the {\it AGASA} data, but not the {\it HiRes} data 
(see Fig. 1)\cite{st04}),
one can also look for the absence of a ``pileup'' spectral feature and for the
absence of photomeson neutrinos, but these may be more difficult to detect.

\section*{Acknowledgments}

Part of this work was supported by NASA grant ATP03-0000-0057.

\end{document}